\documentclass[onecolumn]{revtex4}

\usepackage{graphicx}% Include figure files
\usepackage{dcolumn}% Align table columns on decimal point
\usepackage{bm}% bold math

\input epsf

\begin{document}

\title{\Large Validity of Thermodynamical Laws in Dark Energy Filled Universe}

\author{\bf Samarpita Bhattacharya\footnote{samarpita$_{-}$sarbajna@yahoo.co.in} and Ujjal Debnath\footnote{ujjaldebnath@yahoo.com ,
ujjal@iucaa.ernet.in}}

\affiliation{Department of Mathematics, Bengal Engineering and
Science University, Shibpur, Howrah-711 103, India.}

\date{\today}

\begin{abstract}
We have considered the flat FRW model of the universe which is
filled with only dark energy. The general descriptions of first
and second laws of thermodynamics are investigated on the apparent
horizon and event horizon of the universe. We have assumed the
equation of state of three different types of dark energy models.
We have examined the validity of first and second laws of
thermodynamics on apparent and event horizons for these dark
energies. For these dark energy models, it has been found that on
the apparent horizon, first and second laws are always valid. On
the event horizon, the laws are break down for dark energy models
1 and 2. For model 3, first law cannot be satisfied on the event
horizon, but second law may be satisfied at the late stage of the
evolution of the universe and so the validity of second law on the
event horizon depends on the values of the parameters only.
\end{abstract}


\maketitle

\section{\normalsize\bf{Introduction}}

Recent observation [1, 2] of the luminosity of type Ia supernovae
Wilkinson Microwave Anisotropy Probe, Sloan Digital Sky Survey,
etc. indicate an accelerated expansion of the universe and the
surveys of clusters of galaxies show that the density of matter is
very much less than the critical density. This observation leads
to a new type of matter which violate the strong energy condition.
The matter content responsible for such a condition to be
satisfied at a certain stage of evaluation of the universe is
referred to as dark energy [3-6]. Most of the dark energy models
involve one or more scalar fields with various actions and with or
without a scalar field potential [7]. Now, as the observational
data permits us to have a rather time varying equation of state,
there are a bunch of models characterized by different scalar
fields such as a slowly rolling scalar field, K-essence, tachyonic
field, Chaplygin gas, a phantom model [8-12] etc. In a phantom
model, we have the equation of state as $p  = w \rho$, where $w <
- 1$. The simplest type of phantom model is a scalar field having
a potential V($\phi$) with negative kinetic energy [13]. The
condition $w = - 1$ is named as the phantom divide. There are even
models which can smoothly cross this phantom divide [14].
Currently constrained by the distance measurements of the type Ia
supernova, and the current observational data constrain
the range of equation of state as $-1.38<w< -0.82$ [15].\\

The discovery of Hawking radiation of black holes in the
semi-classical description black hole behaves like a black body
and there is emission of thermal radiation. The temperature of a
black hole is proportional to its surface gravity, and the entropy
is also proportional to its surface area [16]. The Hawking
temperature is given as $T=\frac{\kappa}{2\pi}$, where $\kappa$ is
the surface gravity of a black hole and the entropy of a black
hole is $S=\frac{A}{4G}$, where $A$ is its surface area. At the
apparent horizon, the first law of thermodynamics is shown to be
equivalent to Friedmann equations [17] if one takes the Hawking
temperature and the entropy on the apparent horizon and the
generalized second law of thermodynamics (GSLT) is obeyed at the
horizon. If our Universe can be considered as thermodynamical
system [18], the thermodynamical properties of the black hole can
be generalized to space-time enveloped by the apparent horizon or
the event horizon. The thermodynamical properties of the Universe
may be similar to those of the black hole. Several authors [19]
investigated the properties thermodynamical laws for different
dark energy models.\\

In this work, the general descriptions of first law and GSL of
thermodynamics (using Gibb's law) are investigated on the apparent
horizon and event horizon of the universe. Then we investigate the
thermodynamical properties of an accelerated expanding universe
driven by dark energy with the variable equation of states of the
form $p=w(z)\rho$. Recently, Zhang et al [20] have examined the
validity of first law and generalized second law of thermodynamics
for $w=w_{0}+w_{1}z$. Here we discuss the thermodynamic behavior
by considering another three different parametrizations,
describing the well known dark energy components: (i)
$w(z)=w_{0}+w_{1}\frac{z}{1+z}$~[21]; (ii)
$w(z)=-1+\frac{(1+z)}{3}
\frac{A_{1}+2A_{2}(1+z)}{A_{0}+2A_{1}(1+z)+A_{2}(1+z)^{2}} $~[22]
and (iii) $ w(z)=w_{0}+w_{1}log(1+z)$~[23]. These choices of
$w(z)$ are recently shown to be in good agreement with current
observations in different ranges of $z$. The main motivation of
the present work is that the first law and the GSL of
thermodynamics are valid or not in the dark energy filled universe
which is bounded by the apparent and event horizons. In the three
types of well known dark energy models, we have examined the
validity of the thermodynamical laws in diagrammatically. Finally
we have drawn some interesting conclusions.\\

\section{\normalsize\bf{Validity of Laws of Thermodynamics in Different Dark Energy Models}}

The Einstein field equations for homogeneous, isotropic and flat
FRW universe are given by

\begin{equation}
H^{2}=\frac{8\pi\rho}{3}
\end{equation}
and
\begin{equation}
\dot{H}=-4\pi(\rho+p)
\end{equation}

where $H(=\frac{\dot{a}}{a})$ is the Hubble parameter (choosing
$G=c=1$). The density $\rho$ and pressure $p$ of the dark energy
are connected by the continuity equation

\begin{equation}
\dot{\rho}+3H(\rho+p)=0
\end{equation}

If $w=p/\rho$ be the equation of state of the dark energy then the
above equation can be written as

\begin{equation}
\dot{\rho}+3H(1+w)\rho=0
\end{equation}

Here we'll first present the general discussions of the conditions
of the validity of first law and GSL of thermodynamics on the
apparent and event horizons (which is also presented in our
previous works [24]). Next, we examine the laws are valid or not
on the apparent and event horizons for different dark energy
models.\\

Now we consider the FRW universe as a thermodynamical system with
the horizon surface as a boundary of the system. To study the
generalized second law (GSL) of thermodynamics through the
universe we deduce the expression for normal entropy using the
Gibb's law of thermodynamics [25]

\begin{equation}
T_{X}dS_{I}=pdV+d(E_{X})
\end{equation}

where, $S_{I},~p,~V$ and $E_{X}$ are respectively entropy,
pressure, volume and internal energy within the apparent/event
horizon and $T_{X}$ is the temperature on the apparent horizon
($X=A$)/event horizon ($X=E$). Here the expression for internal
energy can be written as $E_{X}=\rho V$. Now the volume of the
sphere is $V=\frac{4}{3}\pi R_{X}^{3}$, where $R_{X}$ is the
radius of the apparent horizon ($R_{A}$)/event horizon ($R_{E}$)
defined by [25]

\begin{equation}
R_{A}=\frac{1}{H}
\end{equation}
and
\begin{equation}
R_{E}=a\int_{t}^{\infty}\frac{dt}{a}=a\int_{a}^{\infty}\frac{da}{a^{2}H}
\end{equation}

which immediately gives

\begin{equation}
\dot{R}_{E}=HR_{E}-1
\end{equation}

The temperature and the entropy on the apparent/event horizon are

\begin{equation}
T_{X}=\frac{1}{2\pi R_{X}}
\end{equation}
and
\begin{equation}
S_{X}=\pi R_{X}^{2}
\end{equation}

The amount of the energy crossing on the apparent/event horizon is
[26]

\begin{equation}
-dE_{X}=4\pi R_{X}^{3}HT_{\mu\nu}k^{\mu}k^{\nu}dt=4\pi
R_{X}^{3}H(\rho+p)dt=-H\dot{H}R_{X}^{3}dt
\end{equation}

The first law of thermodynamics on the apparent/event horizon is
defined as follows:

\begin{equation}
-dE_{X}=T_{X}dS_{X}
\end{equation}

Rate of change of internal entropy and total entropy are obtained
as

\begin{equation}
\dot{S}_{I}=\frac{\dot{H}R_{X}^{2}(HR_{X}-\dot{R}_{X})}{T_{X}}
\end{equation}
and
\begin{equation}
\dot{S}_{I}+\dot{S}_{X}=2\pi
R_{X}[\dot{H}R_{X}^{2}(HR_{X}-\dot{R}_{X})+\dot{R}_{X}]
\end{equation}

The generalized second law states that total entropy can not be
decreased i.e.,

\begin{equation}
\dot{S}_{I}+\dot{S}_{X}\ge 0
~i.e.,~\dot{H}R_{X}^{2}(HR_{X}-\dot{R}_{X})+\dot{R}_{X} \ge 0
\end{equation}

In our thermodynamic analysis, we are particularly interested in
the parametrizations for the variation of dark energy equation of
state $w$ with redshift $z$ as described below. We discuss the
thermodynamic behavior by considering three different
parametrizations, describing the dark energy component:\\

$\bullet$ {\bf Model ~I:} $w=w_{0}+w_{1}\frac{z}{1+z}$~, where
$w_{0}$ and $w_{1}$ are constants [21] and $z=\frac{1}{a}-1$ is the redshift.\\

In this case, dark energy density can be expressed as (from eq.
(4))
\begin{equation}
\rho=\rho_{0}(1+z)^{3(1+w_{0}+w_{1})}e^{-\frac{3w_{1}z}{1+z}}
\end{equation}

and the field equation (1) reduces to

\begin{equation}
H^{2}=H_{0}^{2}(1+z)^{3(1+w_{0}+w_{1})}e^{-\frac{3w_{1}z}{1+z}}
\end{equation}

where $H_{0}=\sqrt{\frac{8\pi}{3}\rho_{0}}=$ present value of the Hubble parameter (at $a=1$ i.e., $z=0$).\\

From eq. (11), we have the amount of the energy crossing on the
apparent horizon as

\begin{equation}
-dE_{A}=-H\dot{H}R_{A}^{3}dt=-\frac{3}{2}H_{0}^{-1}(1+z)^{-\frac{3}{2}(1+w_{0}+w_{1})}e^{\frac{3}{2}
\frac{w_{1}z}{1+z}}\left(1+w_{0}+\frac{w_{1}z}{1+z}\right)\frac{dz}{1+z}
\end{equation}

The temperature and entropy of the apparent horizon are

\begin{equation}
T_{A}=\frac{H}{2\pi}
\end{equation}
and
\begin{equation}
S_{A}=\pi R_{A}^{2}=\frac{\pi}{H^{2}}
\end{equation}

So we have

\begin{equation}
T_{A}dS_{A}=-\frac{3}{2}H_{0}^{-1}(1+z)^{-\frac{3}{2}(1+w_{0}+w_{1})}
e^{\frac{3}{2}\frac{w_{1}z}{1+z}}\left(1+w_{0}+\frac{w_{1}z}{1+z}\right)\frac{dz}{1+z}
\end{equation}

Thus from equations (18) and (21), we have obtained

\begin{equation}
-dE_{A}=T_{A}dS_{A}
\end{equation}

Therefore in the above model, the first law of thermodynamics is
confirmed on apparent horizon.\\

From equation (14), we get

\begin{equation}
\frac{d(S_{I}+S_{A})}{da}=\frac{d(S_{I}+S_{A})}{dz}\frac{dz}{da}=\frac{9\pi}{2}
H_{0}^{-2}(1+z)^{-4-3(w_{0}+w_{1})}e^{-\frac{3w_{1}z}{1+z}}[(1+w_{0})(1+z)+w_{1}z]^{2}
\geq 0
\end{equation}

So, in this model second law is always valid on the apparent horizon.\\

In this model, the radius of the event horizon in terms of
redshift $z$ can be written as

\begin{equation}
R_{E}=-\frac{1}{1+z}\int_{z}^{-1}\frac{dz}{H}=-\frac{1}{1+z}H_{0}^{-1}\int_{z}^{-1}
(1+z)^{-\frac{3}{2}(w_{0}+w_{1})}e^{\frac{3w_{1}z}{2(1+z)}} dz
\end{equation}

From eq. (11), we have the amount of the energy crossing on the
event horizon as

\begin{equation}
-dE_{E}=\frac{3}{2}R_{E}^{3}H^{2}\left(1+w_{0}+\frac{w_{1}z}{1+z}\right)\frac{dz}{1+z}
\end{equation}

Using equations (9) and (10), we get

\begin{equation}
T_{E}dS_{E}=dR_{E}=\frac{(1-HR_{E})dz}{(1+z)H}
\end{equation}

From (26) and (27), we have seen that $dE_{E}+T_{E}dS_{E}$ is a
function of $z$. In fig.1, we have drawn $dE_{E}+T_{E}dS_{E}$
against redshift $z$ for $w_{0}=-1.2$ and $w_{1}=.98$. It has been
seen that the curve $dE_{E}+T_{E}dS_{E}$ is not coincide with the
$z$ axis, i.e.,

\begin{equation}
dE_{E}+T_{E}dS_{E}\neq 0~,~i.e.,~~  -dE_{E}\neq T_{E}dS_{E}
\end{equation}

So first law does not valid on the event horizon.\\

Now, from eq. (14), we obtain

\begin{equation}
\frac{d(S_{I}+S_{E})}{da}=\frac{d(S_{I}+S_{E})}{dz}\frac{dz}{da}=-(1+z)^{2}\pi\left[2
R_{E}^{4}H\frac{dH}{dz}+
\left\{3R_{E}^{3}H^{2}\left(1+w_{0}+w_{1}\frac{z}{1+z}\right)+2R_{E}\right\}\frac{dR_{E}}{dz}\right]
\end{equation}

which is a complicated function of redshift $z$. In fig.2, we have
drawn $(dS_{I}+dS_{E})$ against redshift $z$ for $w_{0}=-1.2$ and
$w_{1}=.98$. From figure, we have seen that

\begin{equation}
\frac{d(S_{I}+S_{E})}{da}<0
\end{equation}

That means second law of thermodynamics does not hold on the event
horizon of the universe.\\

\begin{figure}
\includegraphics[height=2in]{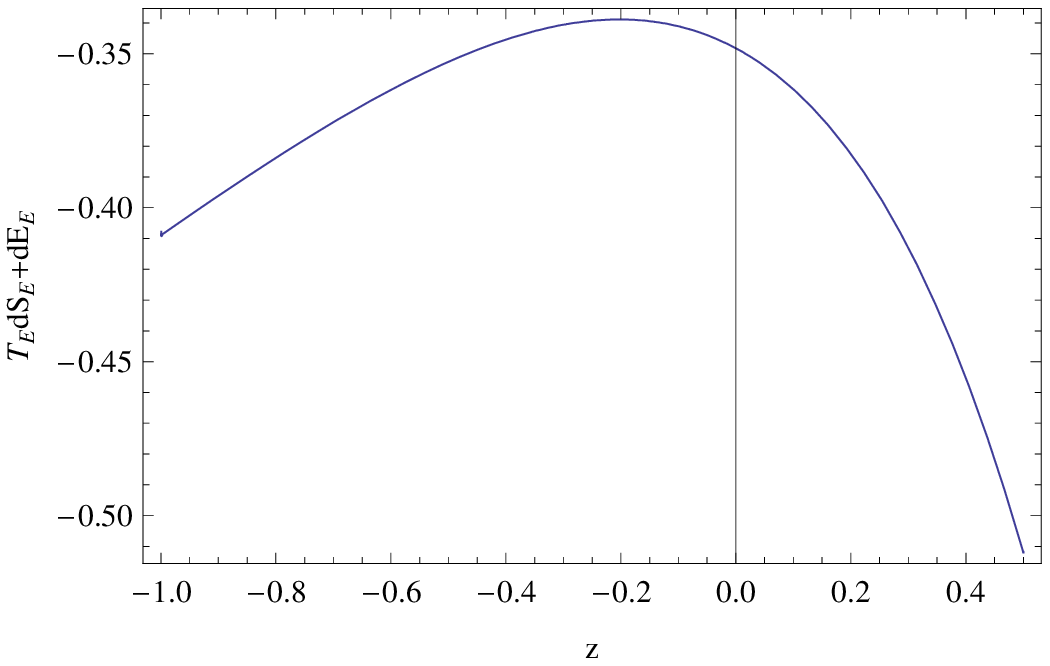}~~~~
\includegraphics[height=2in]{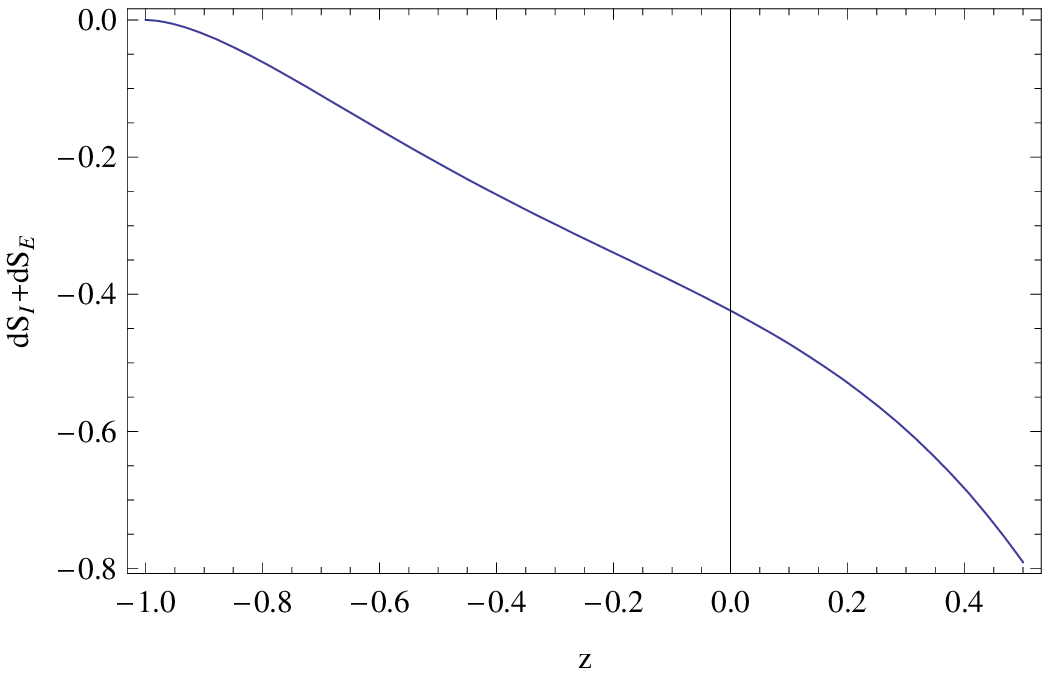}\\
\vspace{1mm} ~~~~~Fig.1~~~~~~~~~~~~~~~~~~~~~~~~~~~~~~~~~~~~~~~~~~~~~~~~~~~~~~~Fig.2\\

\vspace{6mm} Fig. 1 and 2 show the variations of
$(dE_{E}+T_{E}dS_{E})$ and $(dS_{I}+dS_{E})$ respectively against
redshift $z$ for $w_{0}=-1.2$ and $w_{1}=.98$ in model 1.

 \vspace{16mm}

\end{figure}

$\bullet$ {\bf Model 2:} ~~$ w(z)=-1+\frac{(1+z)}{3}
\frac{A_{1}+2A_{2}(1+z)}{A_{0}+2A_{1}(1+z)+A_{2}(1+z)^{2}} $~~,
where $A_{0},~A_{1}$ and $A_{2}$ are constants [22].\\

This ansatz is exactly the cosmological constant $w = -1$ for
$A_{1} = A_{2} = 0$ and DE models with $w =-2/3$ for $A_{0} =
A_{2} = 0$ and $w = -1/3$ for $A_{0} = A_{1} = 0$. It has also
been found to give excellent results for DE models in which the
equation of state varies with time including quintessence,
Chaplygin gas, etc.\\

In this model, equation (1) can be expressed in the form:

\begin{equation}
H^{2}=\frac{H_{0}^{2}}{A_{0}+A_{1}+A_{2}}[A_{0}+A_{1}(1+z)+A_{2}(1+z)^{2}]
\end{equation}

Using equations (9) - (11), we have (on apparent horizon)

\begin{equation}
-dE_{A}=-\frac{1}{2}H_{0}^{-1}
\sqrt{A_{0}+A_{1}+A_{2}}~\frac{[A_{1}+2A_{2}(1+z)]}{\left[A_{0}+A_{1}(1+z)+A_{2}(1+z)^{2}\right]^{\frac{3}{2}}}~dz
\end{equation}
and
\begin{equation}
T_{A}dS_{A}=dR_{A}=-\frac{1}{2}H_{0}^{-1}
\sqrt{A_{0}+A_{1}+A_{2}}~\frac{[A_{1}+2A_{2}(1+z)]}{\left[A_{0}+A_{1}(1+z)+A_{2}(1+z)^{2}\right]^{\frac{3}{2}}}~dz
\end{equation}

From above expressions, we see that

\begin{equation}
-dE_{A}=T_{A}dS_{A}
\end{equation}

Therefore, first law of thermodynamics is satisfied on apparent
horizon of the universe.\\

From equation (14), we get (on apparent horizon)

\begin{equation}
\frac{d(S_{I}+S_{A})}{da}=-\pi H_{0}^{-2}(1+z)^{2}\frac
{(A_{0}+A_{1}+A_{2})[A_{1}+2A_{2}(1+z)]}{\left[A_{0}+A_{1}(1+z)+A_{2}(1+z)^{2}\right]^{2}}
\left[\frac{[A_{1}+2A_{2}(1+z)][2A_{0}+A_{1}(1+z)]}{2[A_{0}+A_{1}(1+z)+A_{2}(1+z)^{2}]}-1\right]
\geq 0
\end{equation}

So, on the apparent horizon, second law is satisfied.\\

Using equations (9) - (11), we have (on event horizon)

\begin{equation}
-dE_{E}=4\pi
R_{E}^{3}H\rho(1+w)dt=-\frac{3}{2}R_{E}^{3}H^{2}\frac{(1+w)dz}{1+z}
\end{equation}
and
\begin{equation}
T_{E}dS_{E}=dR_{E}=-\frac{1}{H_{0}}
\frac{\sqrt{A_{0}+A_{1}+A_{2}}}{\sqrt{A_{0}+A_{1}(1+z)+A_{2}(1+z)^{2}}}dz
\end{equation}

\begin{figure}
\includegraphics[height=2in]{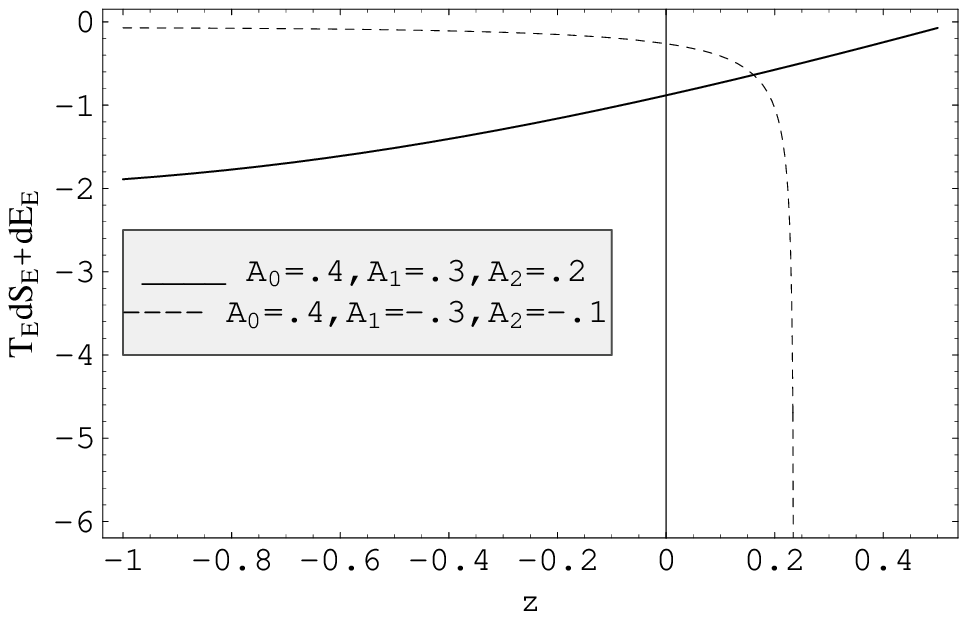}~~~~
\includegraphics[height=2in]{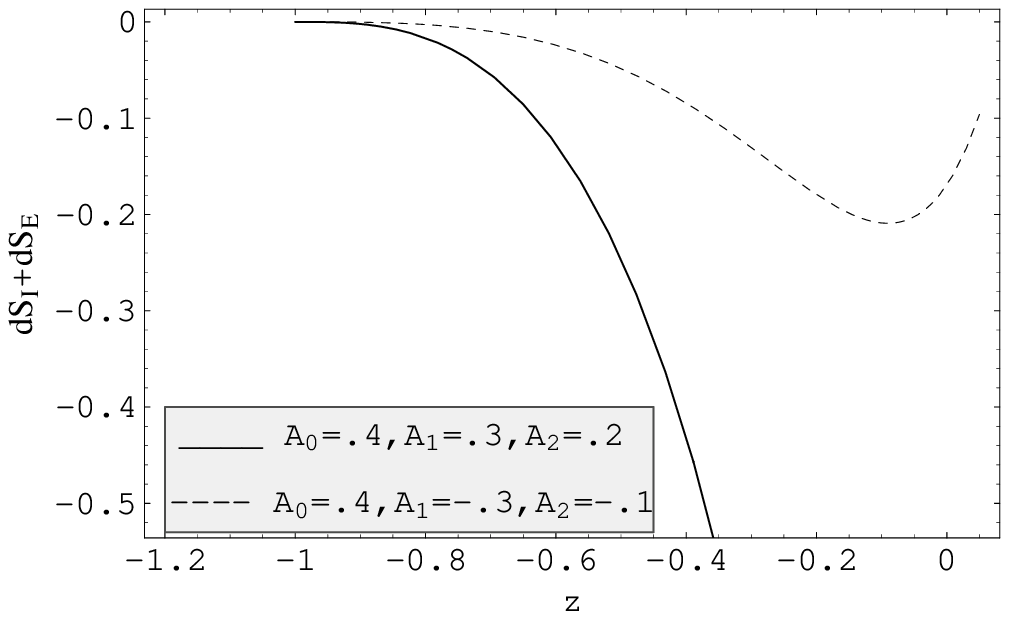}\\
\vspace{1mm} ~~~~~Fig.3~~~~~~~~~~~~~~~~~~~~~~~~~~~~~~~~~~~~~~~~~~~~~~~~~~~~~~~~~~~~~~~~~~~~~~Fig.4\\

\vspace{6mm} Fig. 3 and 4 show the variations of
$(dE_{E}+T_{E}dS_{E})$ and $(dS_{I}+dS_{E})$ respectively against
redshift $z$ for different values of parameters of model 2.

 \vspace{16mm}

\end{figure}

From (34) and (35), we have seen that $dE_{E}+T_{E}dS_{E}$ is a
function of $z$. In fig.3, we have drawn $dE_{E}+T_{E}dS_{E}$
against redshift $z$ for different values of parameters (like,
$A_{0}=.4,A_{1}=.2,A_{2}=.1$ and $A_{0}=.4,A_{1}=-.2,A_{2}=-.1$).
It has been seen that the curve $dE_{E}+T_{E}dS_{E}$ is not
coincide with the $z$ axis, i.e.,

\begin{equation}
dE_{E}+T_{E}dS_{E}\neq 0~,~i.e.,~~  -dE_{E}\neq T_{E}dS_{E}
\end{equation}

that is, first law does not valid on the event horizon. From
equation (14), we get (on event horizon)

\begin{equation}
\frac{d(S_{I}+S_{E})}{da}=-(1+z)^{2}\pi
\left[2R_{E}^{4}H\frac{dH}{dz}+
\left(\frac{R_{E}^{3}H^{2}(1+z)(A_{1}+2A_{2}(1+z))}{(A_{0}+A_{1}(1+z)+A_{2}(1+z)^{2})}+
2R_{E}\right)\frac{dR_E}{dz}\right]
\end{equation}

which is a complicated function of $z$. In fig.4, we have drawn
$(dS_{I}+dS_{E})$ against redshift $z$ for different values of the
parameters like, $A_{0}=.4,A_{1}=.2,A_{2}=.1$ and
$A_{0}=.4,A_{1}=-.2,A_{2}=-.1$. In all the cases, we have seen
that

\begin{equation}
\frac{d(S_{I}+S_{E})}{da}<0
\end{equation}

That means second law of thermodynamics does not hold on the event
horizon of the universe.\\

$\bullet$ {\bf Model 3:} $ w=w_{0}+w_{1}log(1+z) $~, where $w_{0}$
and $w_{1}$ are constants [23].\\

From equation (1), we get

\begin{equation}
H^{2}=H_{0}^{2}(1+z)^{3(1+w_{0})}e^{\frac{3}{2}w_{1}[log(1+z)]^{2}}
\end{equation}

Using equations (9) - (11), we have (on apparent horizon)

\begin{equation}
-dE_{A}=4\pi R_{A}^{3}H\rho(1+w)dt
=-\frac{3}{2}H_{0}^{-1}[1+w_{0}+w_{1}log(1+z)](1+z)^{-\frac{3}{2}(1+w_{0})-1}e^{-\frac{3}{4}w_{1}[log(1+z)]^{2}}~dz
\end{equation}
and
\begin{equation}
T_{A}dS_{A}=dR_{A}=-\frac{3}{2}H_{0}^{-1}[1+w_{0}+w_{1}log(1+z)]
(1+z)^{-\frac{3}{2}(1+w_{0})-1}e^{-\frac{3}{4}w_{1}[log(1+z)]^{2}}~dz
\end{equation}

Thus,

\begin{equation}
-dE_{A}=T_{A}dS_{A}
\end{equation}

So, first law of thermodynamics is valid on the apparent
horizon.\\

From equation (14), we get (on apparent horizon)

\begin{equation}
\frac{d(S_{I}+S_{A})}{da}=\frac{d(S_{I}+S_{A})}{dz}\frac{dz}{da}=
\frac{9}{2}H_{0}^{-2}[1+w_{0}+w_{1}log(1+z)]^{2}(1+z)^{-3(1+w_{0})+1}
e^{-\frac{3}{2}w_{1}[log(1+z)]^{2}} \geq 0
\end{equation}

Therefore, the second  law is still valid on the apparent horizon.\\

The radius of the event horizon in terms of redshift $z$ can be
written as

\begin{equation}
R_{E}=H_{0}^{-1}\left[e^{-\frac{3}{4}w_{1}[log(1+z)]^{2}}(1+z)^{-\frac{3}{2}(1+w_{0})-1}
\left(-\frac{2}{1+3w_{0}}+\frac{2z}{2+3(1+w_{0})}\right)\right]
\end{equation}

Using equations (9) - (11), we have (on event horizon)

\begin{equation}
-dE_{E}=4\pi
R_{E}^{3}H\rho(1+w)dt=-\frac{3}{2}R_{E}^{3}[1+w_{0}+w_{1}log(1+z)]\frac{H^{2}}{1+z}~dz
\end{equation}
and
\begin{equation}
T_{E}dS_{E}=dR_{E}
\end{equation}

\begin{figure}
\includegraphics[height=2in]{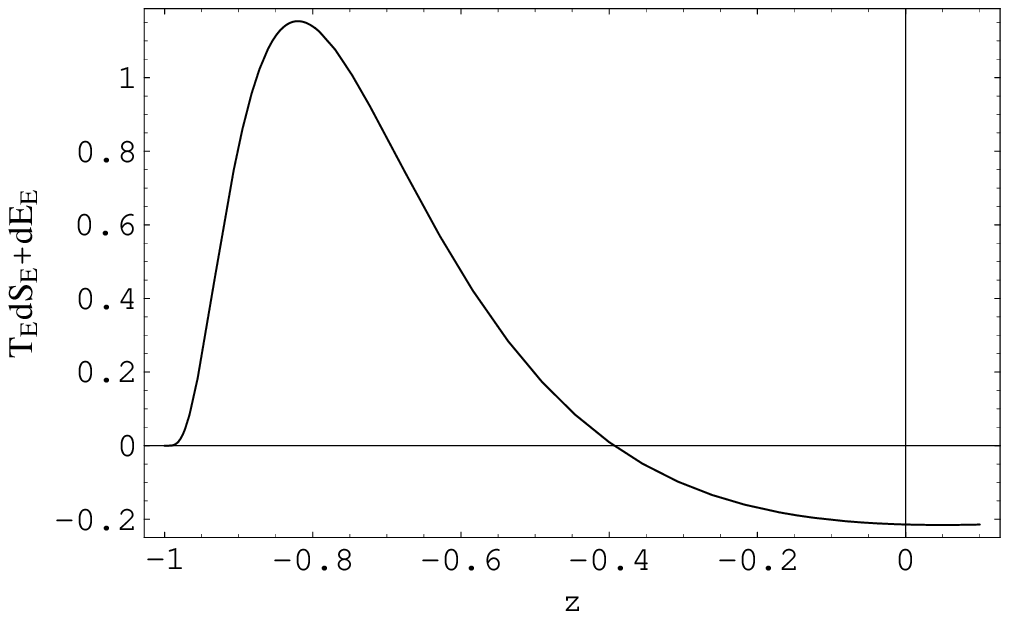}~~~~
\includegraphics[height=2in]{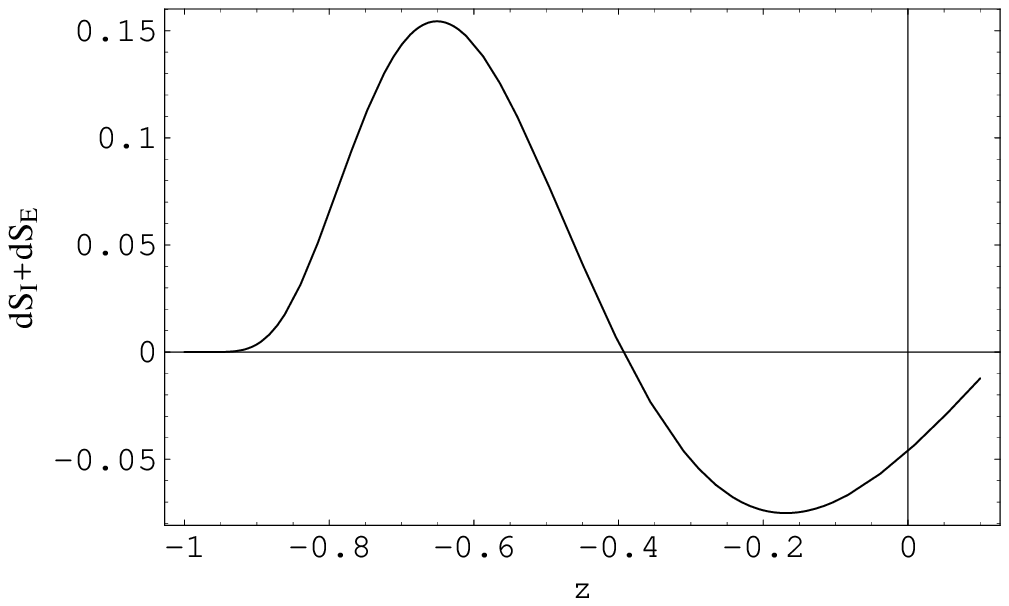}\\
\vspace{1mm} ~~~~~Fig.5~~~~~~~~~~~~~~~~~~~~~~~~~~~~~~~~~~~~~~~~~~~~~~~~~~~~~~~Fig.6\\

\vspace{6mm} Fig. 5 and 6 show the variations of
$(dE_{E}+T_{E}dS_{E})$ and $(dS_{I}+dS_{E})$ respectively against
redshift $z$ for $w_{0}=-0.5$ and $w_{1}=1$ in model 3.

 \vspace{16mm}

\end{figure}

From (45) and (46), we have seen that $dE_{E}+T_{E}dS_{E}$ is a
complicated function of $z$. In fig.5, we have drawn
$dE_{E}+T_{E}dS_{E}$ against redshift $z$. It has been seen that
the curve $dE_{E}+T_{E}dS_{E}$ is not coincide with the $z$ axis,
i.e.,

\begin{equation}
dE_{E}+T_{E}dS_{E}\neq 0~,~i.e.,~~  -dE_{E}\neq T_{E}dS_{E}
\end{equation}

So the first law does not hold on the event horizon.\\

From equation (14), we get (on event horizon)

\begin{equation}
\frac{d(S_{I}+S_{E})}{da}=-(1+z)^{2}\pi\left[2
R_{E}^{4}H\frac{dH}{dz}+\left\{
2R_{E}+3R_{E}^{2}H^{2}(1+w_{0}+w_{1}log(1+z))
\right\}\frac{dR_{E}}{dz} \right]
\end{equation}

In fig.6, we have drawn $(dS_{I}+dS_{E})$ against redshift $z$ for
$w_{0}=-0.5$ and $w_{1}=1$. For these values of the parameters,
the dark energy propagates from quintessence dominated era to
phantom era. From graphical representation, it has been seen that
$\frac{d(S_{I}+S_{E})}{da}$ changes sign from negative to positive
as $z$ decreases. So it may be conclude that in quintessence
dominated era second law does not valid on the event horizon but
in phantom dominated era second law is valid on the event horizon.\\

\section{\normalsize\bf{Discussions}}

We have considered the flat FRW model of the universe which is
filled with only dark energy obeys equation of state $p=w\rho$.
Here $w$ is not a constant, it is assumed as a function of
redshift $z$. The apparent horizon of the universe always exists
and the thermodynamical properties related to the apparent horizon
has been studied. We have investigated to an accelerated expanding
universe driven by DE of time-dependent (i.e., redshift) equation
of state. The event horizon and apparent horizon are in general
different surfaces. The general descriptions of first and second
laws of thermodynamics are investigated on the apparent horizon
and event horizon of the universe. We have assumed the equation of
state of different types of dark energy models. Here we discuss
the thermodynamic behavior by considering three different
parametrizations, describing the dark energy component: (1) Model
1: $w=w_{0}+w_{1}\frac{z}{1+z}$; (2) Model 2: $
w(z)=-1+\frac{(1+z)}{3}
\frac{A_{1}+2A_{2}(1+z)}{A_{0}+2A_{1}(1+z)+A_{2}(1+z)^{2}} $; (3)
Model 3: $ w=w_{0}+w_{1}log(1+z)$. These choices of $w$ are
recently shown to be in good agreement with current observations
in different ranges of $z$.\\

In this work, we have tried to apply the usual definition of the
temperature and entropy as that of the apparent horizon to the
cosmological event horizon and examine the validity of first and
the second laws of thermodynamics. For these dark energy models,
it has been found analytically that on the apparent horizon, first
and second laws are always valid. On the event horizon, we have
found complicated expressions for these dark energy models. So
analytically, we cannot draw any conclusions i.e., it is not
possible to infer about the validity of the thermodynamical laws
on the event horizon for these dark energy models. So graphical
investigations have been needed to draw the final conclusions.
From diagrams, we have been seen that on the event horizon the
laws are break down for dark energy models 1 and 2. For model 3,
first law cannot be satisfied on the event horizon and second law
on the event horizon cannot be satisfied in the early stage of the
universe, but it may be satisfied at the late stage of the
evolution of the universe and so the validity of second law on the
event horizon depends on the values of the parameters only.\\

{\bf Acknowledgement:}\\

The authors are thankful to IUCAA, Pune, India for warm
hospitality where part of the work was carried out.\\

{\bf References:}\\
\\
$[1]$ N. A. Bachall, J. P. Ostriker, S. Perlmutter and P. J. Steinhardt,
\textit{Science} \textbf{284} 1481 (1999); S. J. Perlmutter et al,
\textit{Astrophys. J.} \textbf{517} 565 (1999).\\
$[2]$ A. G. Riess et al., {\it Astron. J.} {\bf 116} 1009 (1998);
A. G. Riess et al., {\it Astrophys. J.} {\bf 607} 665 (2004); C.
L. Bennett et al., {\it Astrophys. J. Suppl. Ser.} {\bf 148} 1
(2003); D. N. Spergel et al., {\it Astrophys. J. Suppl. Ser.} {\bf
170} 377 (2007); D. J. Eisenstein et al., {\it Astrophys. J.} {\bf
633} 560 (2005).\\
$[3]$ V. Sahni and A. A. Starobinsky, {\it Int. J. Mod. Phys. A}
{\bf 9} 373 (2000).\\
$[4]$ P. J. E. Peebles and B. Ratra, {\it Rev. Mod. Phys.} {\bf
75} 559 (2003).\\
$[5]$ T. Padmanabhan, {\it Phys. Rept.} {\bf 380} 235 (2003).\\
$[6]$ E. J. Copeland, M. Sami, S. Tsujikawa, {\it Int. J. Mod.
Phys. D} {\bf  15} 1753 (2006).\\
$[7]$ I. Maor and R. Brustein, {\it Phys. Rev. D} {\bf 67} 103508
(2003); V. H. Cardenas and S. D. Campo, {\it Phys. Rev. D} {\bf
69} 083508 (2004); P. G. Ferreira and M. Joyce, {\it Phys.
Rev. D} {\bf 58} 023503 (1998).\\
$[8]$ B. Ratra and P. J. E. Peebles, {\it Phys. Rev. D} {\bf
37} 3406 (1988).\\
$[9]$ T. Chiba, T. Okabe and M. Yamaguchi, {\it Phys. Rev. D}
{\bf 62} 023511 (2000).\\
$[10]$ A. Sen, {\it JHEP} {\bf 0204} 048 (2002).\\
$[11]$ A. Kamenshchik, U. Moschella and V. Pasquier, {\it Phys.
Lett. B} {\bf 511} 265 (2001); V. Gorini, A. Kamenshchik, U.
Moschella and V. Pasquier, {\it gr-qc}/0403062; V. Gorini, A.
Kamenshchik and U. Moschella, {\it Phys. Rev. D} {\bf 67} 063509
(2003); U. Alam, V. Sahni, T. D. Saini and A. A. Starobinsky, {\it
Mon. Not. R. Astron. Soc.} {\bf 344}, 1057 (2003); H. B. Benaoum,
{\it hep-th}/0205140; U. Debnath, A. Banerjee and S. Chakraborty,
{\it Class.
Quantum Grav.} {\bf 21} 5609 (2004).\\
$[12]$ R. R. Caldwell, {\it Phys. Lett. B} {\bf 545} 23 (2002).\\
$[13]$ S. M. Carroll, M. Hoffman and M. Trodden, {\it Phys. Rev.
D} {\bf 68} 023509 (2003).\\
$[14]$ X. Meng, M. Hu and J. Ren, {\it astro-ph}/0510357.\\
$[15]$ A. Melchiorri, L. Mersini and M. Trodden, {\it Phys. Rev.
D} {\bf 68} 043509 (2003).\\
$[16]$ S. W. Hawking, {\it Commun. Math. Phys.} {\bf 43} 199
(1975); J. D. Bekenstein, {\it Phys. Rev. D} {\bf 7} 2333
(1973).\\
$[17]$ R. G. Cai and S. P. Kim, {\it JHEP} {\bf 02} 050 (2005).\\
$[18]$ G. W. Gibbons and S. W. Hawking, {\it Phys. Rev. D} {\bf
15} 2738 (1977).\\
$[19]$ B. Wang, Y. G. Gong and E. Abdalla, {\it
Phys. Rev. D} {\bf 74} 083520 (2006); Y. Gong, B. Wang and A.
Wang, {\it JCAP} {\bf 01} 024 (2007); M. R. Setare, {\it Phys.
Lett. B} {\bf 641} 130 (2006); Y. Zhang, Z. Yi, T. -J. Zhang and
W. Liu, {\it Phys. Rev. D}
{\bf 77} 023502 (2008).\\
$[20]$ Y. Zhang, Z. Yi, T. -J. Zhang and W. Liu, {\it Phys. Rev. D} {\bf 77} 023502 (2008).\\
$[21]$ E. V. Linder, {\it Phys. Rev. Lett.} {\bf 90} 091301
(2003); T. Padmanabhan and \
T. R. Choudury, {\it Mon. Not. R. Astron. Soc.} {\bf 344} 823 (2003); A. A. Sen, {\it JCAP} {\bf 0603} 010 (2006).\\
$[22]$ U. Alam, V. Sahni, T. D. Saini and A. A. Starobinski, {\it
Mon. Not. R. Astron. Soc.} {\bf 354} 275 (2004); U. Alam, V.
Sahni and A. A. Starobinski, {\it JCAP} {\bf 0406} 008 (2004).\\
$[23]$ G. Efstathiou, {\it Mon. Not. R. Astron. Soc.} {\bf 310}
842 (1999); R. Silva, J. S. Alcaniz and J. A. S. Lima, {\it Int.
J. Mod. Phys. D} {\bf 16} 469 (2007).\\
$[24]$ S. Bhattacharya and U. Debnath, arXiv: 1006.2600[gr-qc];
arXiv: 1006.2609[gr-qc].\\
$[25]$ B. Wang, Y. G. Gong and E. Abdalla, {\it Phys. Rev. D} {\bf
74} 083520 (2006); G. Izquierdo and D. Pavon, {\it Phys. Lett. B}
{\bf 633} 420 (2006).\\
$[26]$ R. Bousso, {\it Phys. Rev. D} {\bf 71} 064024 (2005).\\

\end{document}